\documentclass[twoside]{ilcws07}
\usepackage[latin1]{inputenc}
\usepackage[dvips]{graphicx,epsfig,color}
\usepackage{wrapfig,rotating}
\usepackage{amssymb,amsmath,array}

\begin{document}
\title{
Polarised Geant4 -- Applications at the ILC} 
\author{Andreas Sch{\"a}licke$^1$, Karim Laihem$^2$ and Pavel Starovoitov$^3$
\vspace{.3cm}\\
1- DESY \\
Platanenallee 6, 15378 Zeuthen - Germany
\vspace{.1cm}\\
2- RWTH Aachen - Phys. Inst. IIIB \\
Physikzentrum, 52056 Aachen- Germany
\vspace{.1cm}\\
3- NCPHEP BSU\\
M.Bogdanovich str.\ 153, 220040 Minsk - Belarus\\
}


\maketitle
\vspace{-6cm}
\hfill DESY-07-202\\[-6mm]
\vspace{6cm}

\begin{abstract}

Geant4 is a Monte Carlo simulation framework for the description of
interactions of particles and matter. 
Starting with version 8.2 a new package of QED physics processes is
available, allowing for the studies of interactions of polarised
particles with polarised media dedicated to beam applications. 
In this contribution some details about the implementation are
presented and applications to the linear collider are
discussed.  
\end{abstract}

\section{Introduction}
Programs that can simulate the complex interaction patterns of
particles traversing matter are indispensable tools for the design and
optimisation of particle detectors. A major example of such programs
is Geant4 \cite{Agostinelli:2002hh,geant4}, which is widely used in
high energy physics, medicine, and 
space science. Different parts of this tool kit can be combined to
optimally fulfil the users needs. A powerful geometry package allows
the creation of complex detector configurations. The physics
performance is based on a huge list of interaction processes. Tracking
of particles is possible in arbitrary electromagnetic fields.
However, polarisation has played only a minor role so
far\footnote{Compton scattering of linearly polarised photons is
  available since Geant4 version 3.1. Polarised Rayleigh scattering and
  Photoelectric effect of linearly polarised photons have been
  addressed recently.}. 

The new extension in the library of electromagnetic physics is
dedicated to polarisation effects in beam applications 
\cite{Dollan:2005nj}.
It aims for a proper treatment of longitudinal polarised
electrons/positrons or circularly polarised photons and their
interactions with polarised matter. 

Polarised versions of Bhabha/M{\o}ller scattering (B/MS), electron-positron
annihilation (EPA), Compton scattering (CS), pair creation (PC), and
bremsstrahlung (BS) are already part of the polarisation library.
A polarised version of the Photoelectric Effect is in preparation.

Two basic problem classes are addressed: 
\begin{itemize}
\item 
   {\em Polarisation transfer}
   from initial beam particles to secondaries created in material
   interactions can be investigated. For instance, in the context of
   the ILC positron source a detailed study of the production
   mechanism of polarised positrons from photons emitted from a helical
   undulator is now possible. 
\item 
   Interactions of polarised particles with polarised matter can
   be simulated. In general, 
   asymmetries may be observed if beam and target particles are
   polarised. 
   They 
   manifest themselves in total 
   cross sections as well as in differential distributions, which
   provides the basis of applications in {\em polarimetry}.
\end{itemize}
These new features have already been exploited in the analysis of
data from the E166 experiment \cite{Alexander:2003fh}, and are also used in
studies for an anticipated {low energy polarimeter} \cite{Schaelicke:2006zz} 
as well as in design and performance optimisation studies for 
an ILC positron source \cite{ushakov:pac07}.

\begin{figure}
\centerline{%
\begin{picture}(340,170)
 \put(0,0){\includegraphics[width=6cm]{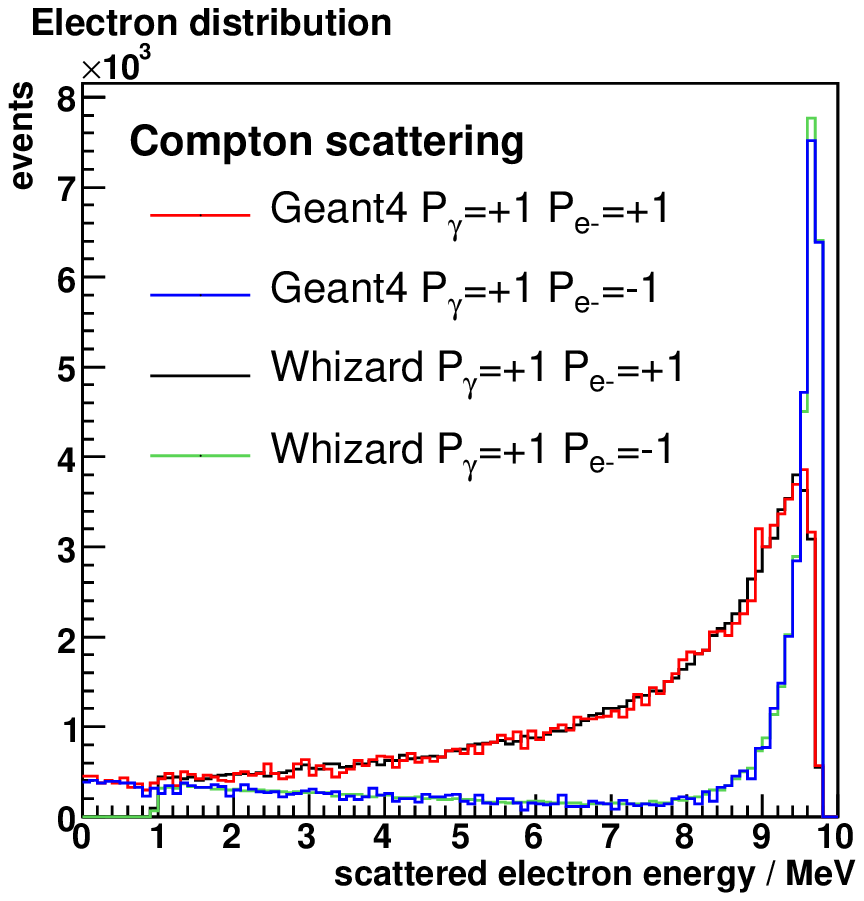}}
 \put(170,0){\includegraphics[width=6cm]{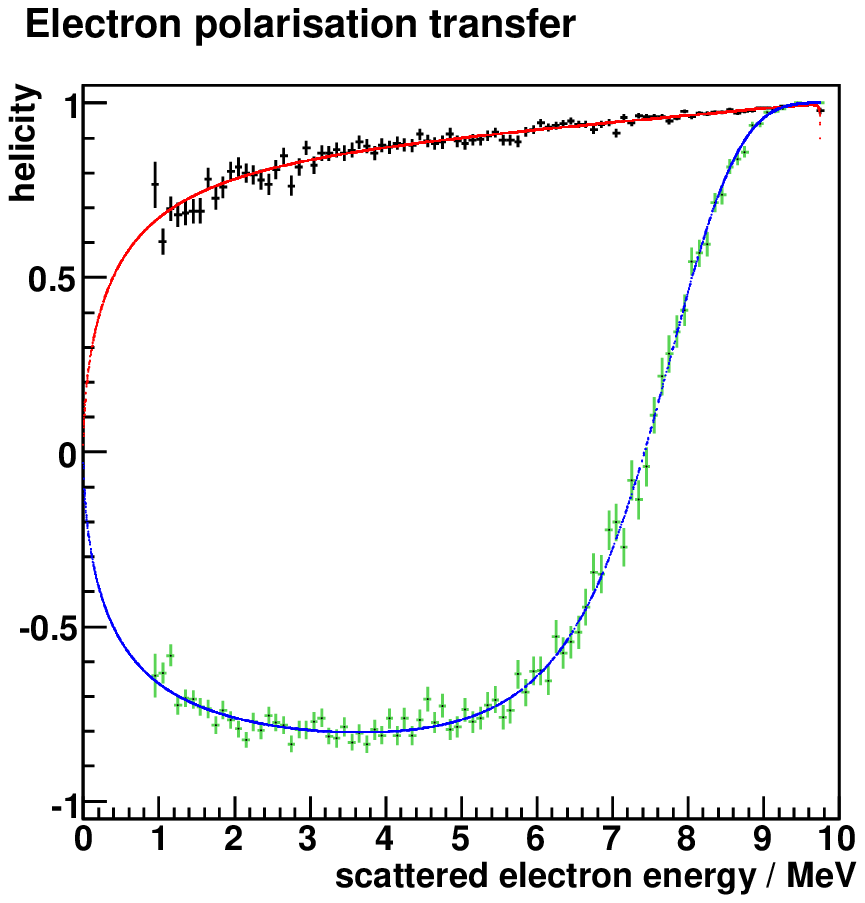}}
\end{picture}%
}
\caption{Compton scattering. Distribution (left) of electrons and
  their degree of polarisation (right) are compared with results from
  Whizard/O'mega for different initial state helicities \cite{omega}.} 
\label{Fig:comptonDists}
\end{figure}
\section{The new polarisation library}

Several simulation packages for the realistic description
of the evolution of electromagnetic showers in matter have been
developed. A prominent example is EGS (Electron Gamma 
Shower)\cite{Nelson:1985ec}. 
For this simulation framework, extensions with the treatment of
polarised particles exist \cite{Namito:1993sv,Liu:2000ey}; 
the most complete has been developed by 
\cite{Floettmann:thesis}. 
It is based on the matrix formalism
\cite{McMaster:1961}, which enables a very general treatment of
polarisation. However, the Fl{\"o}ttmann extension concentrates on
evaluation of polarisation transfer, i.e.\ the effects of polarisation
induced asymmetries are neglected, and interactions with polarised
media are not considered.  
Another important simulation tool for detector studies is \textsc{Geant3}
\cite{Brun:1985ps}. Here also some effort has been made to include
polarisation \cite{Alexander:2003fh,Hoogduin:thesis}, but these
extensions are not publicly available.

In general, the implementation of polarisation in the
library in Geant4 follows very closely the approach by
\cite{McMaster:1961}. 
A {\em Stokes vector} is associated to each particle and used  
to track the 
polarisation from one interaction to another. 


Five new process classes for CS, BS, B/MS, PC, EPA with polarisation
are now available for physics studies with Geant4. The implementation
has been carefully checked against existing references, 
alternative codes, and dedicated analytic calculations. 
Figure \ref{Fig:comptonDists} shows exemplarily a comparison of electron
distribution and polarisation transfer in Compton scattering using
Whizard/O'mega\footnote{Whizard/O'mega is a multipurpose matrix
  element generator allowing for polarisation in the initial state. It
  is dedicated to high energy collision with many particles 
  in the final state, but can also be used to calculate polarised
  $2\to2$ cross sections, like e.g.\ Compton scattering.} \cite{omega}. 
Further details can be found in
\cite{Laihem:thesis,G4:PRM}. 

\newpage
\begin{wrapfigure}{r}{0.50\columnwidth}
\vspace{-2.cm}
\centerline{\includegraphics[width=0.45\columnwidth]{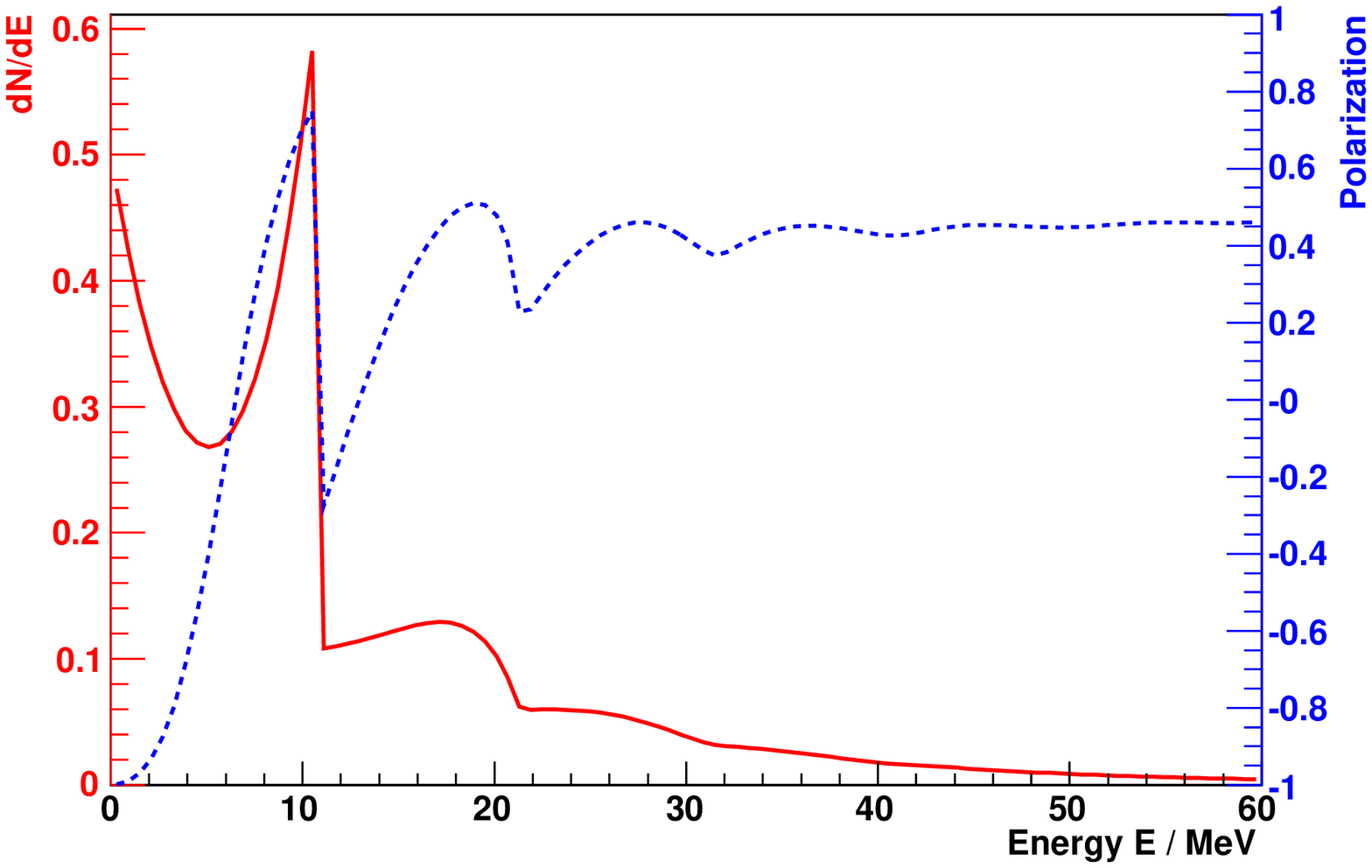}}
\caption{Energy (red) and polarisation (blue) of
  photons created in a helical undulator. 
}
\label{Fig:ilcphotons}
\centerline{\includegraphics[width=0.45\columnwidth]{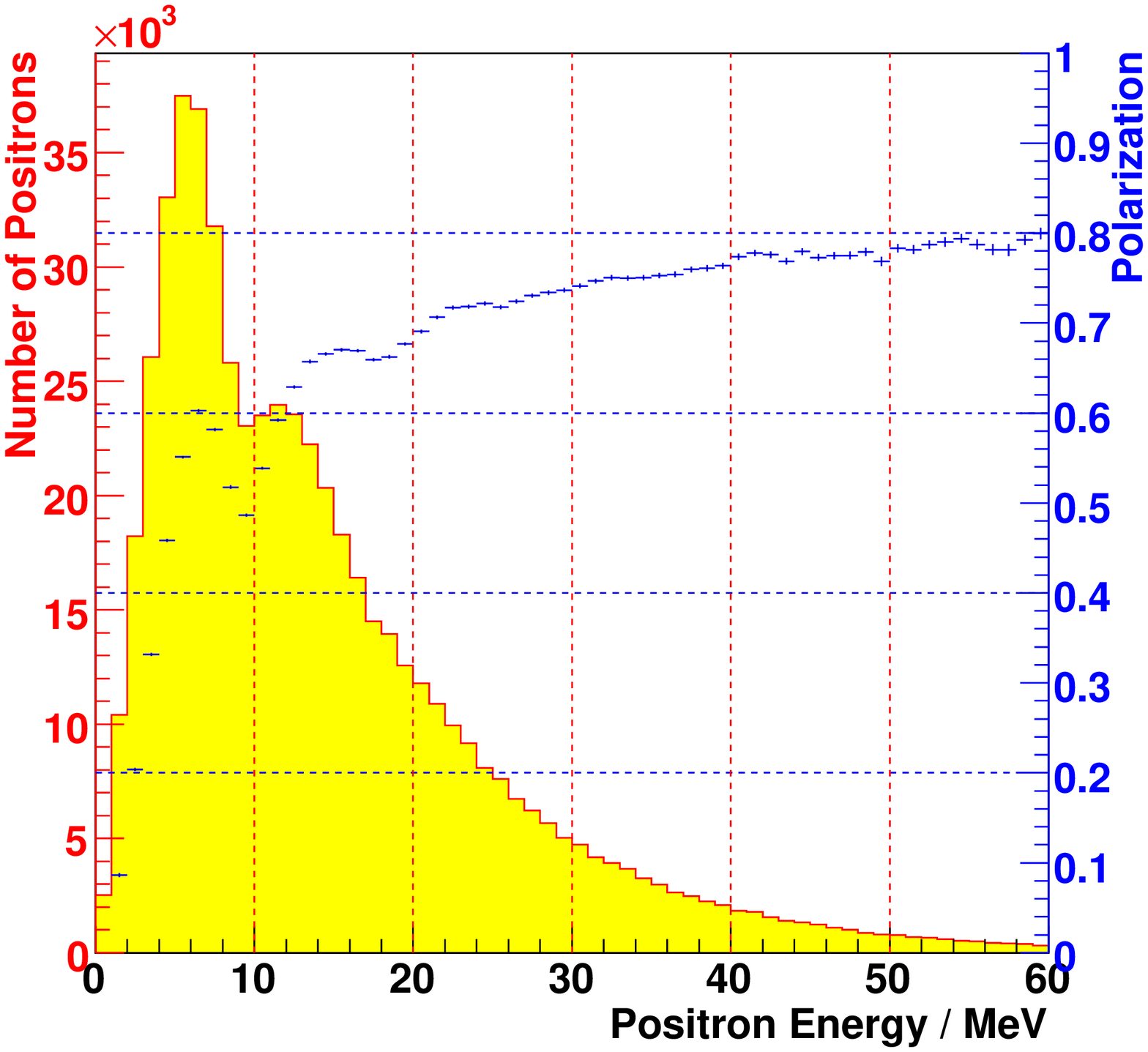}}
\caption{Energy (red) and polarisation (blue) of positrons 
 after the production target of  thickness $d=0.4 X_0$. }
\label{Fig:ilcpositrons}
\vspace{-1.cm}
\end{wrapfigure}
\section{Applications to the ILC}

A key feature of the ILC will be that both beams -- electrons and
positrons -- are polarised. 
With the new polarisation extension it is now possible to
investigate details of the production mechanism and polarimetry
options for electrons and, in particular, for positrons.

The degree of electron and positron
polarisation should be known at least to an accuracy of a few per
mill at the collision point to take full advantage of measurements
with polarised beams.  

\subsection{Polarised positron source}

In the baseline design of the ILC \cite{RDR} polarised positrons are
produced from circularly polarised photons created in an helical
undulator hitting a thin Ti target. 
The spin of the photon is
transferred to the  electron-positron pairs produced resulting in
a net polarisation of the particles emerging from the target.  
The positrons are captured just behind the target in a dedicated
capture optics, i.e.\ an adiabatic matching device, and their degree
of polarisation has to be maintained until they reach the 
collision point. 

Figure \ref{Fig:ilcphotons} pictures energy distribution of photons
and their degree of polarisation as expected from an ideal helical
undulator with strength $K=1$ and period
$\lambda=1\operatorname{cm}$. The resulting positron energy and
polarisation distributions after the production target
are shown in Figure
\ref{Fig:ilcpositrons}. 

\subsection{Low energy polarimeter}

\begin{figure}
\centerline{\includegraphics[width=0.45\columnwidth]{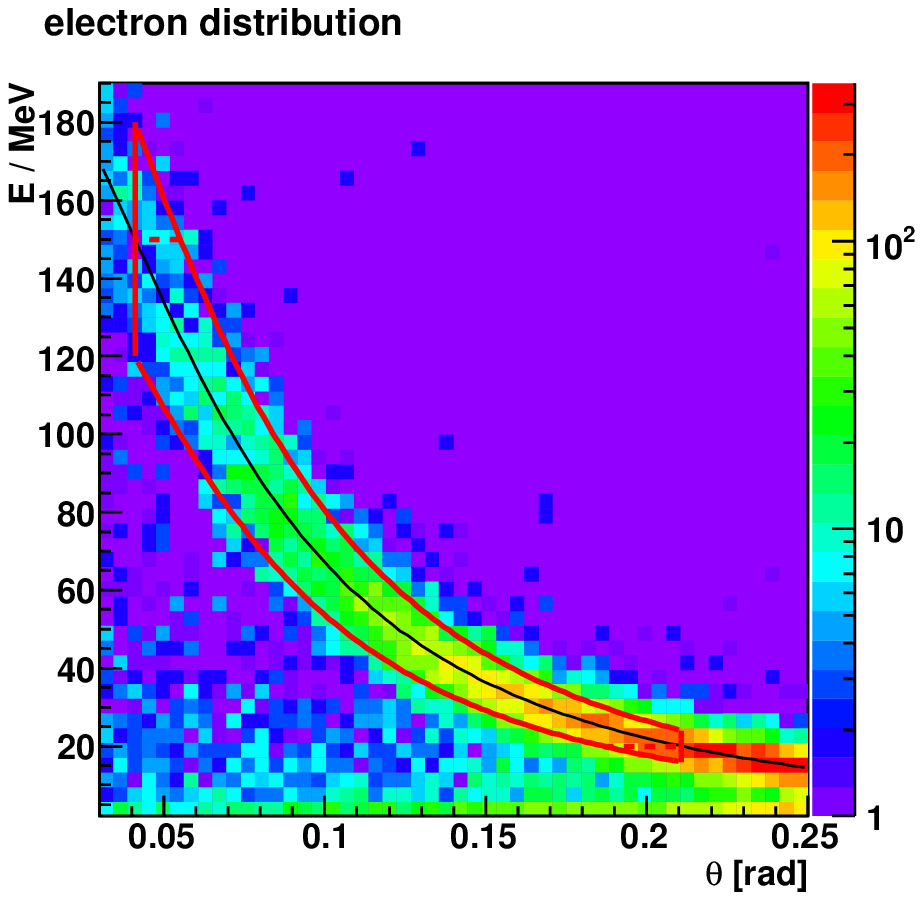}
            \includegraphics[width=0.45\columnwidth]{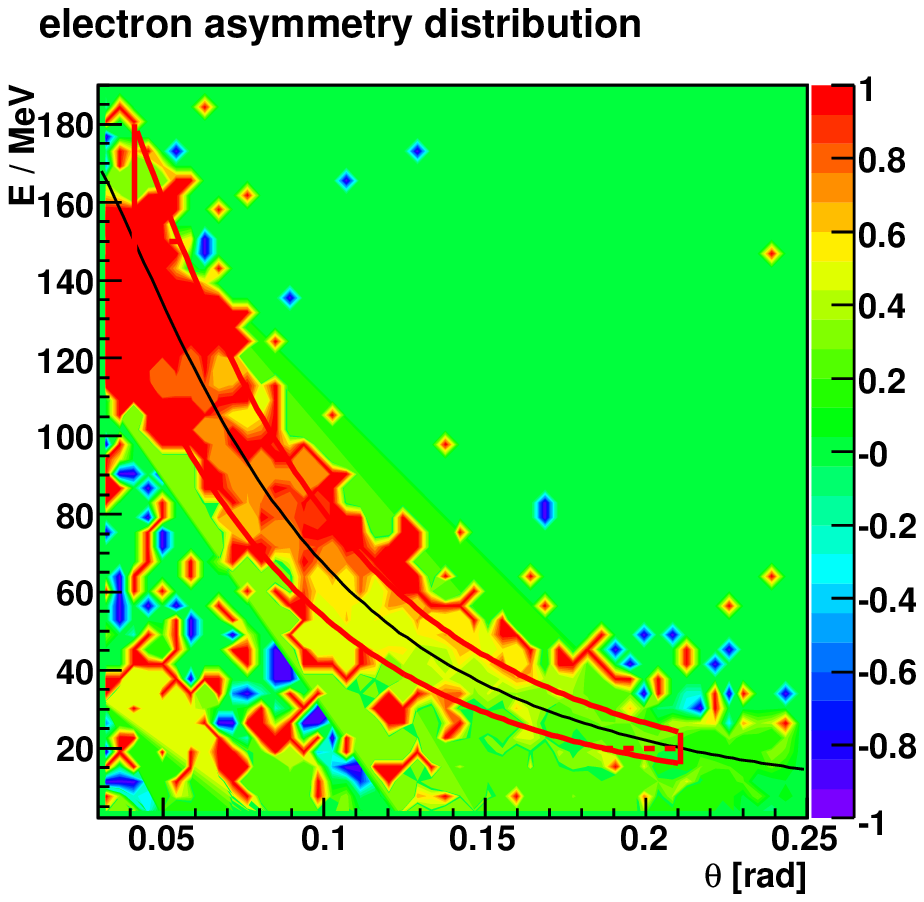}}
\caption{Angular distribution of electrons  (left) and analysing
  power (right). 
} 
\label{Fig:lepolElectrons}
\end{figure}

For commissioning and optimisation of the ILC operation, an independent check of the
polarisation near the creation point of positrons is recommended.
A Bhabha polarimeter \cite{Schaelicke:2006zz} 
is a promising candidate to realise a low energy positron polarisation
measurement.
There, a thin magnetised iron foil (few 10$\mu$m
think) is placed in the positron beam. A few of the positrons hitting
the foil undergo Bhabha scattering. Rate and distribution of the
scattered electrons and positrons depend on the polarisation of the
beam, and can be exploited for polarimetry. The dominating background
are bremsstrahlung positrons, which can be substantially reduced by
looking at the electron distribution only.

The left part of Figure \ref{Fig:lepolElectrons} shows an
energy vs.\ angle distribution of electrons emerging from 
a 30$\mu$m iron foil hit by a beam with $2\cdot10^{10}$ positrons
of 200 MeV. 
The right part of Figure
\ref{Fig:lepolElectrons} gives the corresponding analysing power. In the
central acceptance approx.\ $10^4$ electrons per positron bunch are
expected with an analysing power of about 40\%.

\subsection{The E166 experiment}

A proof-of-principle experiment has been carried out at SLAC to
demonstrate the production of polarised positrons in a manner suitable for
implementation at the ILC \cite{Alexander:2003fh}. A 
helical undulator of 2.54 mm period and 1 m length produced circularly
polarised photons, with a first harmonic endpoint energy of 8 MeV, when
traversed by a 46.6 GeV electron beam. The polarised photons were
converted to polarised positrons in a 0.2-radiation-length tungsten
target. The polarisation of these positrons was measured at several
energies using a Compton transmission polarimeter. 

Geant4 simulations using the polarisation extension have been
employed in the determination of the expected polarisation
profile. These simulations also provided the basis for the
determination of the analysing power needed
to determine the polarisation 
of the produced positron beam. Further details may be found in
\cite{url,Laihem:thesis,Star:2007}. 

\section{Summary} 

Starting with version 8.2 a new package of QED physics processes has
been added to the Geant4 framework, allowing studies of polarised
particle interactions with polarised media. Applications include
design and optimisation of a polarised positron source and beam
polarimetry for a future linear collider facility.  

\section*{Acknowledgements}

The authors are indebted to A.\ Stahl as the initiator of this
project, and also would like to thank T.~Lohse and S.~Riemann for
fruitful collaboration, and helpful discussions. 

%


\begin{footnotesize}



%

\end{footnotesize}


\end{document}